\documentclass{aa}
\usepackage{txfonts}
\usepackage{graphicx}
\usepackage{natbib}
\usepackage{color}
\definecolor{DarkBlue}{rgb}{0.00,0.00,0.55}
\usepackage{hyperref}
\hypersetup{
 colorlinks=true,
 citecolor=DarkBlue,
 linkcolor=DarkBlue,
 urlcolor=DarkBlue}

\begin{document}

\title{Small-scale magnetic flux emergence in a sunspot light bridge}

\author{Rohan E. Louis\inst{1,}\inst{2,}\inst{3},
  Luis R. Bellot Rubio\inst{2},
  J. de la Cruz Rodr\'{\i}guez\inst{4}, 
  H. Socas-Navarro\inst{5,}\inst{6},
  \and A. Ortiz\inst{2}}

\authorrunning{Louis et al.}  
\titlerunning{Small-scale magnetic flux
  emergence in a sunspot light bridge}

\offprints{Rohan E. Louis $[$rlouis@aip.de$]$}

\institute{Leibniz-Institut f\"ur Astrophysik Potsdam (AIP),
	  16 An der Sternwarte, 14482 Potsdam, Germany
          \and Instituto de Astrof{\'\i}sica de Andaluc{\'\i}a (CSIC),
	  Apartado de Correos 3004, 18080 Granada, Spain
          \and Udaipur Solar Observatory, Physical Research Laboratory,
          Dewali, Badi Road, Udaipur -313004, Rajasthan India       
          \and Institute for Solar Physics, Dept. of Astronomy,  Stockholm University, Albanova University Center,  SE-10691 Stockholm, Sweden
          \and Instituto de Astrof\'{\i}sica de Canarias,
          V\'{\i}a L\'actea s/n, 38205-La Laguna, Tenerife, Spain
          \and Departamento de Astrof\'\i sica, Universidad de La Laguna, 38205,
  La Laguna, Tenerife, Spain
}

\date{Received 29 June 2015 / Accepted 19 August 2015}

\abstract
{Light bridges are convective intrusions in sunspots that often show
  enhanced chromospheric activity.}
   {We seek to determine the nature of flux emergence in a light bridge and the processes related to its evolution in the solar atmosphere.}
{We analyse a sequence of high-resolution spectropolarimetric
  observations of a sunspot taken at the Swedish 1-m Solar Telescope.
  The data consist of spectral scans of the photospheric Fe~{\sc i}
  line pair at 630 nm and the chromospheric \ion{Ca}{ii}~854.2~nm
  line. Bisectors were used to construct Dopplergrams from the
  \ion{Fe}{i} 630.15 nm measurements. We employed LTE and non-LTE 
  inversions to derive maps of physical parameters in the photosphere
  and chromosphere, respectively.}
{We observe the onset of blueshifts of about 2~km~s$^{-1}$ near the
  entrance of a granular light bridge on the limbward side of the
  spot. The blueshifts lie immediately next to a strongly redshifted
  patch that appeared six~minutes earlier. Both patches can be seen for
  25~minutes until the end of the sequence. The blueshifts coincide
  with an elongated emerging granule, while the redshifts appear at
  the end of the granule. In the photosphere, the development of the
  blueshifts is accompanied by a simultaneous increase in field
  strength of about 400~G. The field inclination increases by some
  25$^\circ$, becoming nearly horizontal. At the
  position of the redshifts, the magnetic field is equally horizontal
  but of opposite polarity.  An intense brightening is seen in the
  \ion{Ca}{ii} filtergrams over the blueshifts and
    redshifts, about 17 minutes after their detection in the
  photosphere. The brightening is due to emission in the blue wing of
  the Ca~{\sc ii}~854.2 nm line, close to its knee.  Non-LTE
  inversions reveal that this kind of asymmetric emission is caused by a
  temperature enhancement of $\sim$700~K between $-5.0 \le \log\tau
  \le -3.0$ and a blueshift of 3~km~s$^{-1}$ at $\log\tau=-2.3$ that
  decreases to zero at $\log\tau=-6.0$ }
{The photospheric blueshifts and redshifts observed in a granular
  light bridge seem to be caused by the emergence of a small-scale,
  flat $\Omega$-loop with highly inclined footpoints of opposite
  polarity that brings new magnetic field to the surface.  The gas
  motions detected in the two footpoints are reminiscent of a siphon
  flow. The rising loop is probably confined to the lower atmosphere
  by the overlying sunspot magnetic field and the interaction between
  the two flux systems may be responsible for temperature enhancements
  in the upper photosphere/lower chromosphere. This is the first time
  that magnetic flux is observed to emerge in the strongly magnetised
  environment of sunspots, pushed upwards by the
  convective flows of a granular light bridge.}

\maketitle

\section{Introduction}
\label{intro}
Light bridges (LBs) are bright structures in the dark umbra of
sunspots \citep{1979SoPh...61..297M}, which bear an umbral, penumbral, or
quiet Sun morphology depending on their lifetimes
\citep{1994ApJ...426..404S, 2004SoPh..221...65L, 2008ApJ...672..684R,
  2012ApJ...755...16L}. They are often seen during the formation and
decay of sunspots along fissures where individual fragment spots
coalesce or split \citep{1987SoPh..112...49G}. These structures are
conceived to be either ``field-free" intrusions of plasma in the umbral
magnetic field \citep{1979ApJ...234..333P, 1986ApJ...302..809C,
  2006A&A...447..343S} or signatures of magneto-convection
\citep{2004ApJ...604..906R}.

Many LBs show a central dark lane running parallel to their axis and
tiny bright grains or granules \citep{1994ApJ...426..404S,
  2002A&A...383..275H, 2003ApJ...589L.117B, 2004SoPh..221...65L,
  2008ApJ...672..684R}. The dark lane appears to be the result of a
hot upflowing plume braked by the surrounding magnetic field, which
forces the plume into a cusp-like shape
\citep{2006ApJ...641L..73S,2010ApJ...720..233C}. Upflows of nearly 500
m~s$^{-1}$ are sometimes seen along the central dark lane, flanked by
less vigorous downflows \citep{2010ApJ...718L..78R}.
\citet{2008A&A...489..747G} reported weaker upflows of about
70~m~s$^{-1}$ along the dark lane and downflows of 150~m~s$^{-1}$
sideways.

The magnetic field in LBs is weak and inclined
\citep{1997ApJ...484..900L, 2006A&A...453.1079J, 2007PASJ...59S.577K},
possibly because of the magnetic canopy formed by the adjacent umbral
field over the LB \citep{2006A&A...453.1079J}.  This complex and
dynamic magnetic configuration is believed to be responsible for a
wide range of chromospheric and coronal activity, including surges in
H$\alpha$ \citep{2001ApJ...555L..65A,2010ApJ...711.1057G}, C-class
flares \citep{2003ApJ...589L.117B}, coronal jet ejections
\citep{2012PASA...29..193L}, and persistent chromospheric brightenings
\citep{2008SoPh..252...43L, 2014A&A...567A..96L, 
2009ApJ...696L..66S}. Some of these
brightenings are co-spatial with supersonic downflows in the
photosphere \citep{2009ApJ...704L..29L}. This kind of activity has been
attributed to magnetic reconnection \citep{2012ASPC..454..209M},
although other mechanisms cannot be ruled out
\citep{2011ApJ...738...83S}.

Our knowledge of the general magnetic structure of LBs is still based
on snapshots that reflect different phases of their evolution. Apart
from imaging observations, which are routinely carried out 
\citep{2002A&A...383..275H, 2007PASJ...59S.577K,2012ApJ...755...16L},
only a handful of spectroscopic or polarimetric time sequences of
LBs have been acquired in the past \citep{2003SoPh..215..261S,
  2008ApJ...672..684R,2012A&A...537A..19R, 2013A&A...560A..84S}. The
evolution of the magnetic and velocity fields on short timescales is
dictated by processes that are ultimately responsible for the physical
nature and dynamism of LBs. Thus it is important to ascertain how the
photospheric magnetic field in LBs influences the overlying
chromosphere at the relevant timescales.  This kind of study demands 
high-resolution spectroscopy and polarimetry in the two atmospheric layers
with good temporal cadence.

Here we report on the emergence of small-scale magnetic flux in a
granular LB using a time sequence of spectropolarimetric observations
taken at the Swedish 1-m Solar Telescope. The response of the
chromosphere to this event is seen as strong brightenings at the
location of the emerging flux nearly 17~minutes after the magnetic
field appears on the photosphere. We describe the observed Stokes
profiles and invert them to investigate the properties of the emerging
flux, its evolution, and its influence on the chromosphere.

\section{Observations and data processing}
\label{data}

On 2009 July 5, the leading sunspot of AR 11024 was observed with the
CRisp Imaging SpectroPolarimeter \citep[CRISP;][]{2008ApJ...689L..69S}
at the Swedish 1-m Solar Telescope \citep{2003SPIE.4853..341S}. CRISP
is a dual Fabry-P\'erot system capable of performing 
quasi-simultaneous measurements in several spectral lines with full
Stokes polarimetry.  On that particular day, we acquired
spectropolarimetric scans of the \ion{Fe}{i} lines at 630.15 and
630.25~nm and the \ion{Ca}{ii} line at 854.2~nm.

Each of the two Fe lines was scanned at 15 wavelength positions in
steps of 4.8~pm, covering a spectral window of $-33.6$ to $+33.6$ pm
from line centre. In addition, we observed a continuum point at
630.320~nm. \ion{Ca}{ii} 854.2~nm was scanned at 17 wavelength
positions from $-80$ to $+80$ pm in steps of 10~pm, plus a continuum
point at $+0.24$~nm. Polarimetry was carried out by modulating the
incoming light with liquid crystal variable retarders.  At each
wavelength position, nine exposures per modulation state were taken,
resulting in a total of 36 exposures. The time needed to scan the two
Fe lines and the continuum point was 33~s.  Changing the prefilter
took another 5~s, followed by 19~s to scan the Ca line.

The observations were reduced using the CRISP data pipeline
\citep{2015A&A...573A..40D} with the telescope model of
\cite{2010arXiv1010.4142S}. High spatial resolution was achieved 
by means of the adaptive optics system of the SST
\citep{2003SPIE.4853..370S} and the Multi-Object Multi-Frame Blind
Deconvolution technique \citep{2005SoPh..228..191V}.

The seeing conditions were good and allowed us to obtain a complete
data set from 15:05 to 16:50~UT, when the sunspot was located at a
heliocentric angle of $\theta = 33^{\circ}$ ($\mu = \cos \theta =
0.84$). The images cover an area of $58 \times 57\arcsec$ with a
spatial sampling of 0\farcs059~pixel$^{-1}$. More details of the
observations can be found in \citet{2012ApJ...757...49W}.

\begin{figure}[t]
\centering
\includegraphics[bb= 0 -612 560 0,angle=90,width = 1.2\columnwidth]{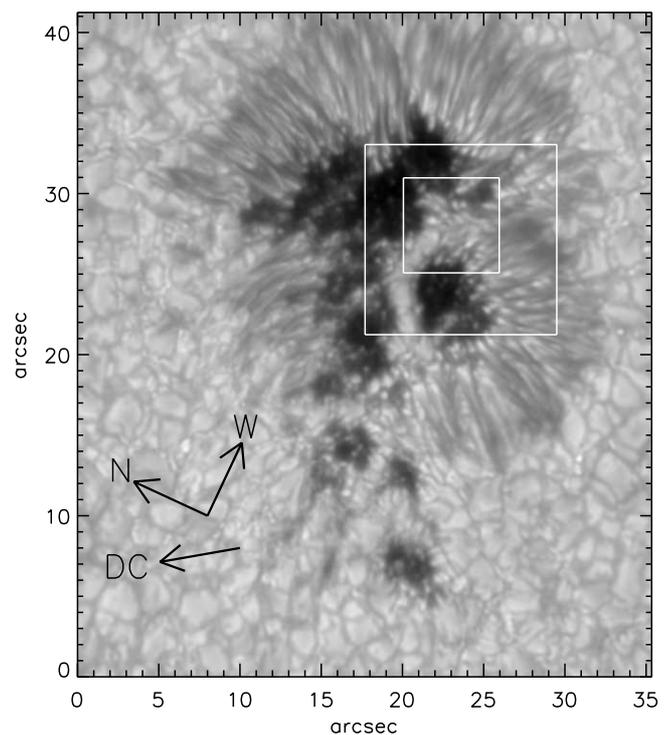}
\caption{Continuum image of the leading sunspot of AR 11024 at 630~nm
  on 2009 July 5, 16:44 UT. The large and small white boxes correspond
  to the field of view shown in Figs.~\ref{figure02} and \ref{figure03},
  respectively.  DC indicates the disc centre position.}
\label{figure01}
\end{figure}

Figure~\ref{figure01} shows a continuum image of the main spot of AR
11024 recorded through the CRISP narrowband channel. A granular LB is
clearly seen, anchored to the eastern and southern penumbra. The spot
shows disruptions in the penumbra near the eastern part of the LB. 
The region chosen for analysis consists of the
LB and a portion of the limbside penumbra (boxed area in the figure).

\section{Results}
\label{res}
\subsection{Strong photospheric flows and chromospheric activity}
\label{flows}

\begin{figure*}[t]
\centering
\includegraphics[angle=90,width = \textwidth]{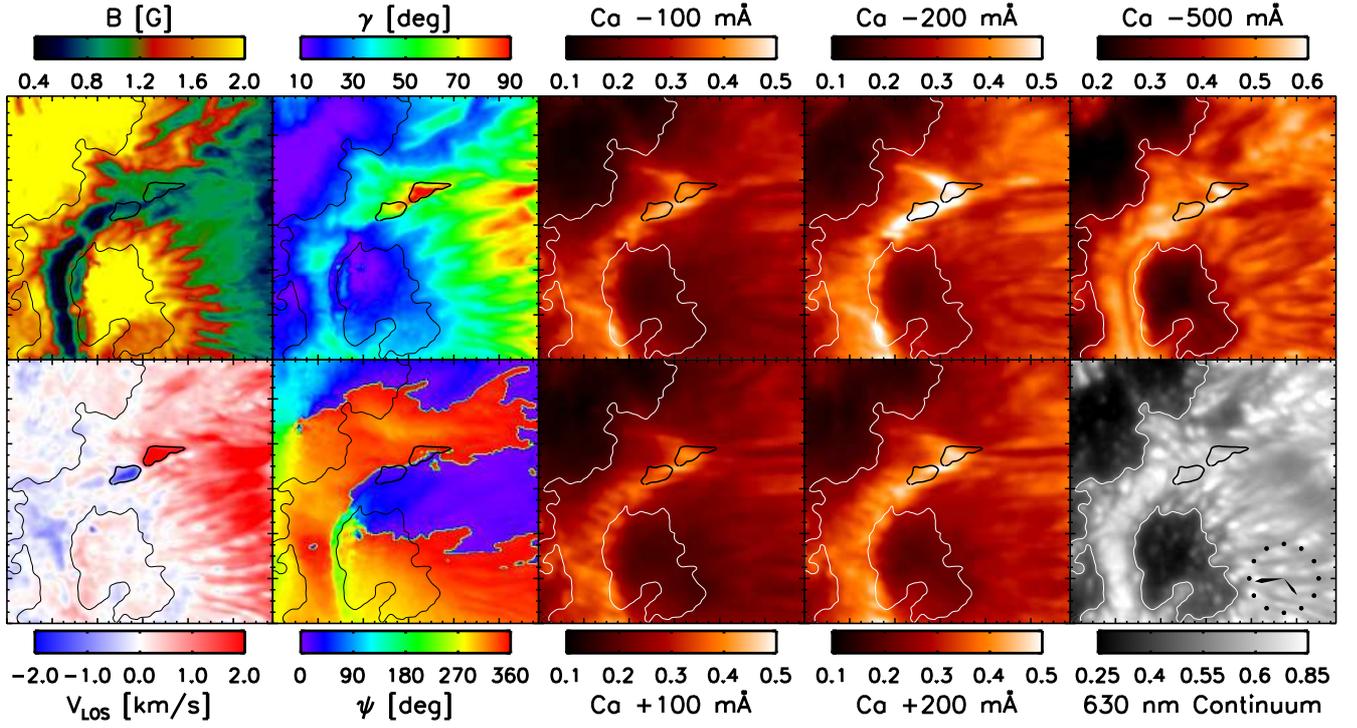}
\caption{Physical parameters and \ion{Ca}{ii} intensity filtergrams of
  the LB and its surroundings. The $12\arcsec\times12\arcsec$ field of view
  corresponds to the large square indicated in Fig.~\ref{figure01}.  
  Column 1: magnetic field strength (top) and Dopplergram from bisectors at the 70\%
  intensity level (bottom).  The field strength map corresponds to an
  optical depth of $\log\tau=-1.0$.  Column 2: field inclination (top)
  and azimuth (bottom) in the local reference frame.  Inclinations of
  0$^\circ$ and 180$^\circ$ imply that the field is directed away
  from and into the solar surface, respectively. Zero
  azimuth is along the positive $x$-axis and increases in the
  counter-clockwise direction. Columns 3--5: intensity filtergrams at
  different wavelength positions of the Ca~{\sc ii}~854.2 nm line and
  the continuum image at 630~nm.  All images have been scaled
  according to their respective colour bars.  The maps correspond to
  scan number 97 acquired at 16:44~UT.  The temporal
  evolution is shown in a movie available in the online edition.}
\label{figure02}
\end{figure*}

Figure~\ref{figure02} presents a photospheric Dopplergram of the LB
and its surroundings at 16:44 UT (first panel, bottom row).
Dopplergrams were computed from bisectors at different intensity
levels of the \ion{Fe}{i} 630.15~nm line using the procedure described
in \citet{2010ApJ...718L..78R}, with the dark umbral regions serving
as a reference for the absolute wavelength scale. The Dopplergram shows
a patch of blueshifts of $\sim$1.5 km~s$^{-1}$ next to a strong
redshift of 3~km~s$^{-1}$ on the entrance of the LB.  Blueshifts in
other parts of the LB are typically 700~m~s$^{-1}$, while the limbside
penumbra displays redshifts ranging from 1 to 2~km~s$^{-1}$ associated
with the Evershed flow \citep{1909MNRAS..69..454E}. The patches of
strong blueshifts and redshifts (hereafter referred to as the blue and
red patches, respectively) have an area of $\sim$0.8 arcsec$^2$ and
are separated by about 1\farcs6.

To determine the magnetic field configuration associated with these
photospheric flows, we inverted the \ion{Fe}{i} 630~nm Stokes profiles
observed in the region of interest with the SIR code \citep[Stokes
Inversion based on Response functions;][]{1992ApJ...398..375R}. A
single magnetic component in each pixel was used, perturbing
temperatures, magnetic field strengths and line-of-sight (LOS)
velocities at two nodes. The field inclination and azimuth were kept
constant with height. We also considered a constant value of
macroturbulence (0.5~km~s$^{-1}$) and zero stray light in all pixels.
The field inclination and azimuth were subsequently transformed to the
local reference frame. They are shown along with the field strength in
Fig.~\ref{figure02}.

The blue and red patches have field strengths of about 820~G and
900~G, respectively, whereas the typical field strength in the rest of
the LB, particularly along its axis, is of the order of 400~G and therefore
much weaker.  The field strength map corresponds to an optical depth
of $\log\tau=-1.0$ since the response of the \ion{Fe}{i}~630 nm lines
to magnetic fields is maximum around that height
\citep{2005A&A...439..687C}.

Figure~\ref{figure02} also shows that the red patch consists of nearly
horizontal fields (inclinations of 90$^\circ$) with the polarity
changing sign in several pixels. These highly inclined fields extend
partially into the blue patch where the mean inclination is 62$^\circ$
but individual pixels reach a maximum of 83$^\circ$. By comparison,
the magnetic field in the arm of the LB is relatively vertical, with
inclinations of 40$^\circ$. The two patches sit close to a section of
the sunspot where the field azimuth on either side confluences by some
40$^\circ$. This arises from the two azimuth centres located in the
two umbral cores partitioned by the LB.

Columns 3--5 in Fig.~\ref{figure02} show intensity filtergrams at
different wavelength positions within the \ion{Ca}{ii}~854.2~nm line.
The blue and red patches are co-spatial with enhanced brightness that
is stronger in the blue wing compared with the red wing, particularly
near the knee of the line.  The brightenings are also observed at
other positions of the line wing (e.g. $-500$~m\AA), but they are more
localised. The conspicuous \textsf{V}-shaped chromospheric brightening
seen at $-200$~m\AA\/ (fourth column, top panel) has an intensity of
0.57$I_{\textrm{\tiny{QS}}}$ and is nearly 20\% brighter than the
neighbourhood.

\subsection{Temporal evolution of the flow structures}
\label{evol}
\begin{figure*}[t]
\centering
\includegraphics[bb=0 0 792 930,width = 1.05\textwidth]{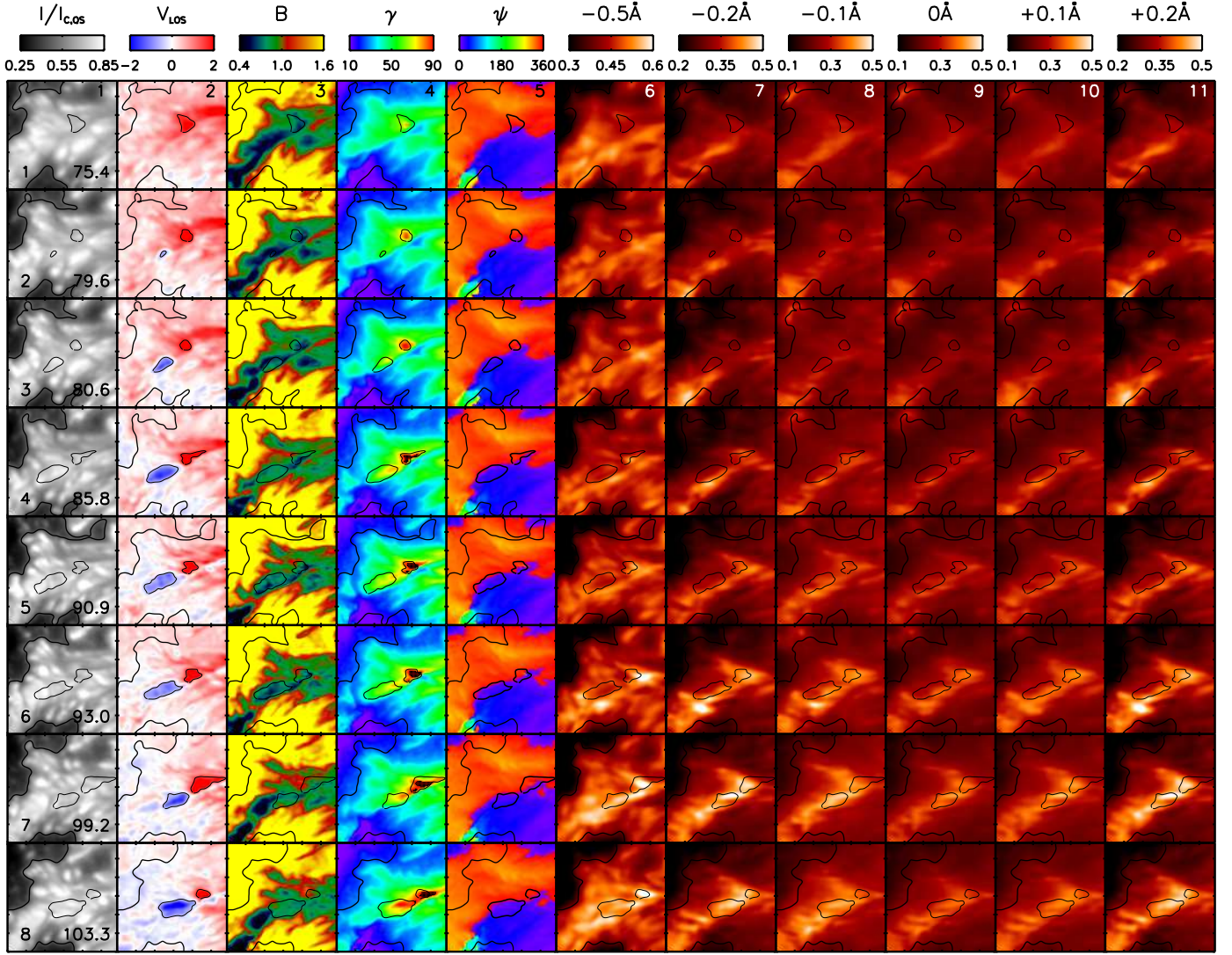}
\vspace{-240pt}
\caption{Evolution of blueshifts and redshifts in the LB and their
  corresponding signature in the chromosphere. The
  $6\arcsec\times6\arcsec$ field of view corresponds to the small
  square indicated in Fig.~\ref{figure01}. Columns 1--5: continuum
  intensity at 630 nm, LOS velocity, magnetic field strength,
  inclination, and azimuth, respectively.  In col. 4, all pixels
  exceeding an inclination of 90$^{\circ}$ are coloured black.
  Columns 6--11: intensity filtergrams at different wavelength
  positions of the Ca~{\sc ii}~854.2 nm line. All images have been
  scaled according to their respective colour bars.  The
    first column gives the row numbers and times of the scan at the
    bottom of each panel. Column numbers are indicated at the top of
    each panel in row 1.}
\label{figure03}
\end{figure*}

Figure~\ref{figure03} shows the development of the blueshifts and
redshifts in the LB and the associated chromospheric brightenings
during the last 30~minutes of observations. The sequence starts at minute~75
(16:20~UT) with the red patch already present two minutes earlier.
The blue patch appears four minutes later and grows significantly from
0.07~arcsec$^2$ to 0.97~arcsec$^2$ in a time span of three minutes,
becoming nearly as large as the red patch.  Both patches remain next
to each other, close to the entrance of the LB, until the end of the
sequence, implying lifetimes greater than 25~minutes.  The blue patch
maintains an elliptical structure with its major axis nearly parallel
to that of the LB. The movie accompanying Fig.~\ref{figure02} 
shows that the blue patch coincides with an elongated emerging 
granule (rows 2 and 3 of Fig.~\ref{figure03}). The red patch lies near the 
end of this granule, which is interrupted by the penumbral filaments. At that
position, the radial arrangement of the penumbra further out is also
disrupted. The two patches do not exhibit significant proper motions,
but evolve in situ with their respective centroids spaced apart by
about 1\farcs6.

\begin{figure*}[t]
\centerline{
\includegraphics[angle=0,width = 0.5\textwidth]{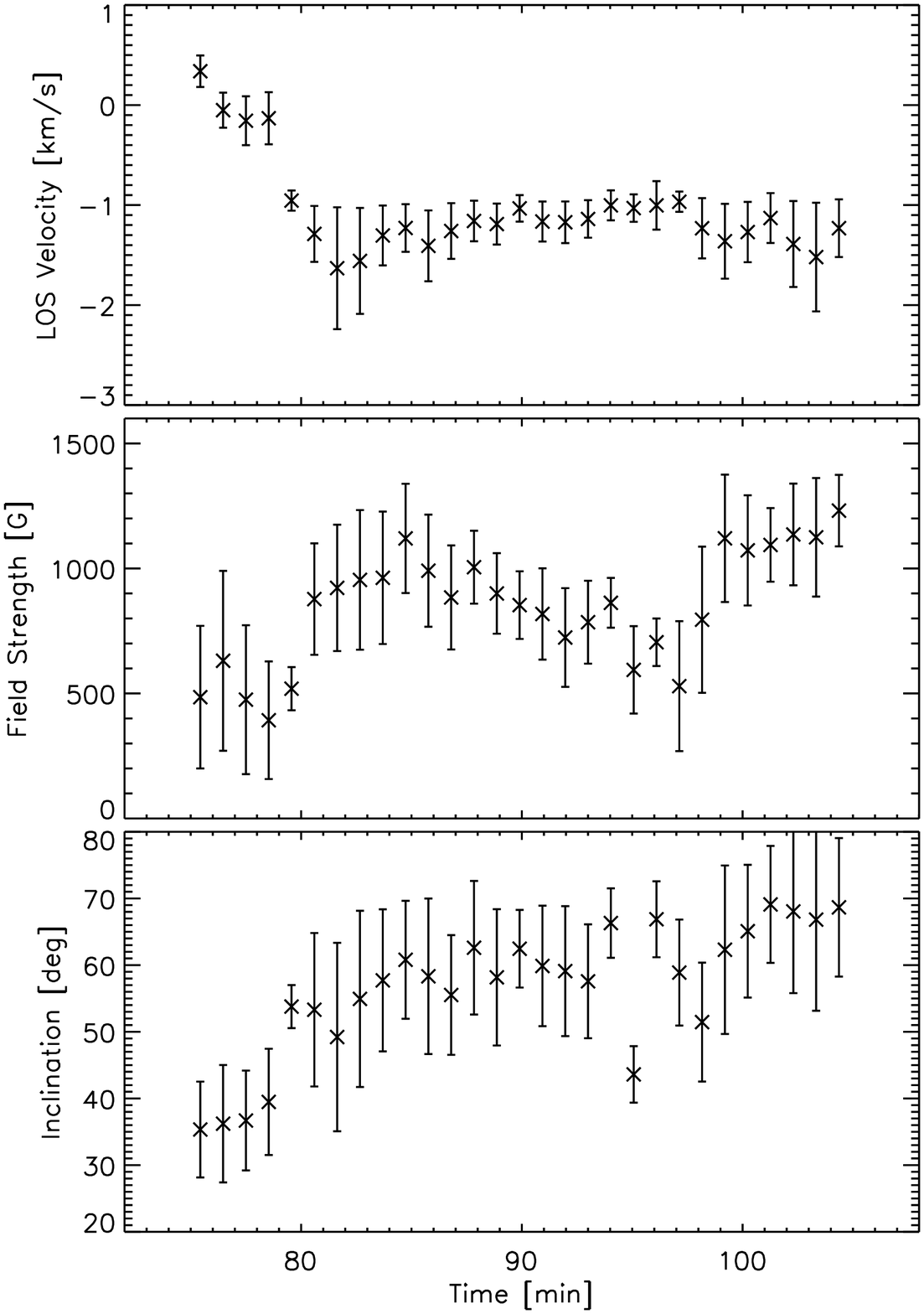}
\includegraphics[angle=0,width = 0.5\textwidth]{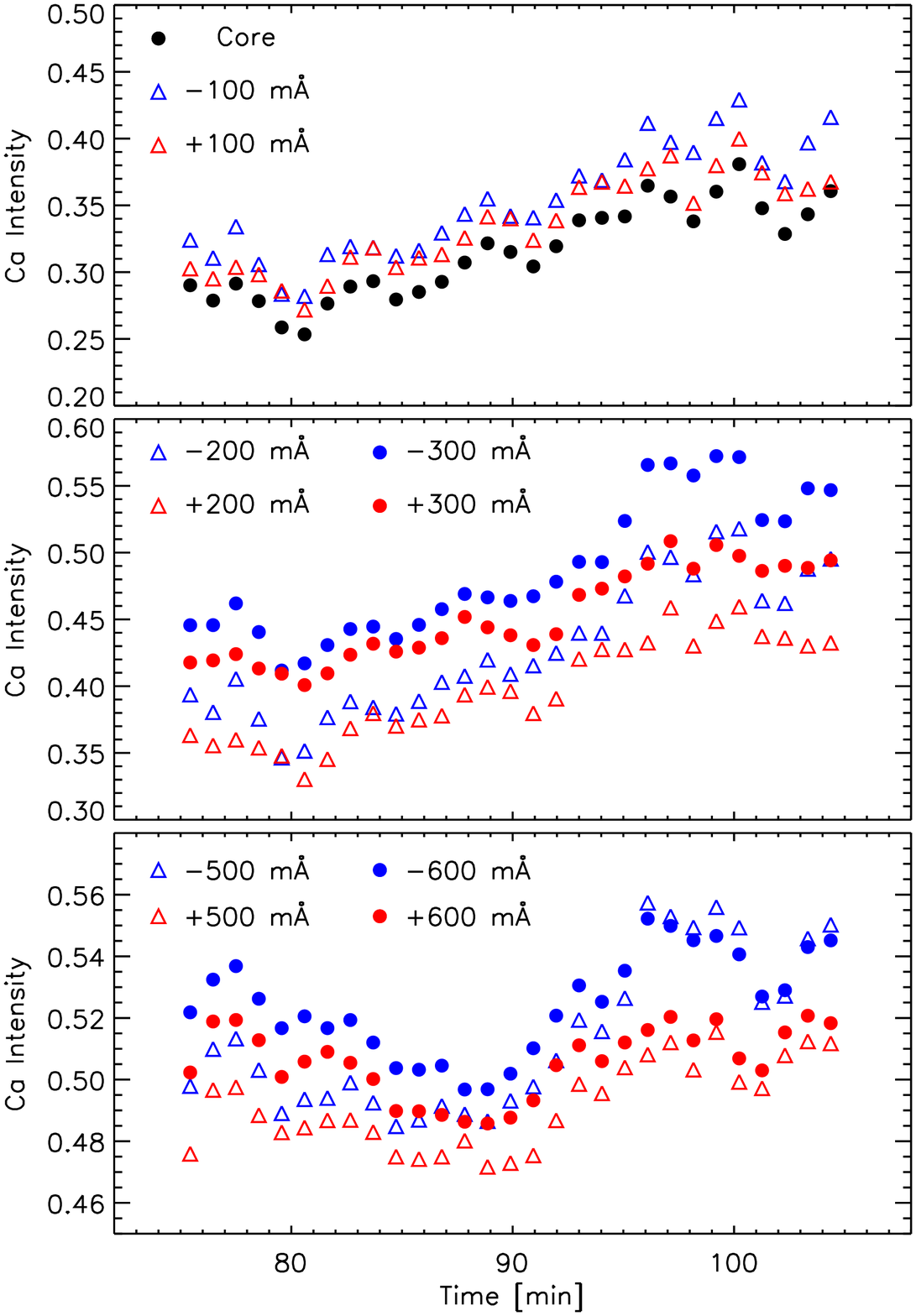}
}
\vspace{-5pt}
\caption{Temporal variation of physical parameters averaged over the
  blue patch. The set of panels on the right correspond to the mean Ca
  intensity at different wavelength positions. The vertical bars in
  the left panels represent the standard deviation of the individual
  values entering the mean.}
\label{figure04}
\end{figure*}

\begin{figure}[!h]
\centering
\includegraphics[bb=20 0 540 792,width = \columnwidth]{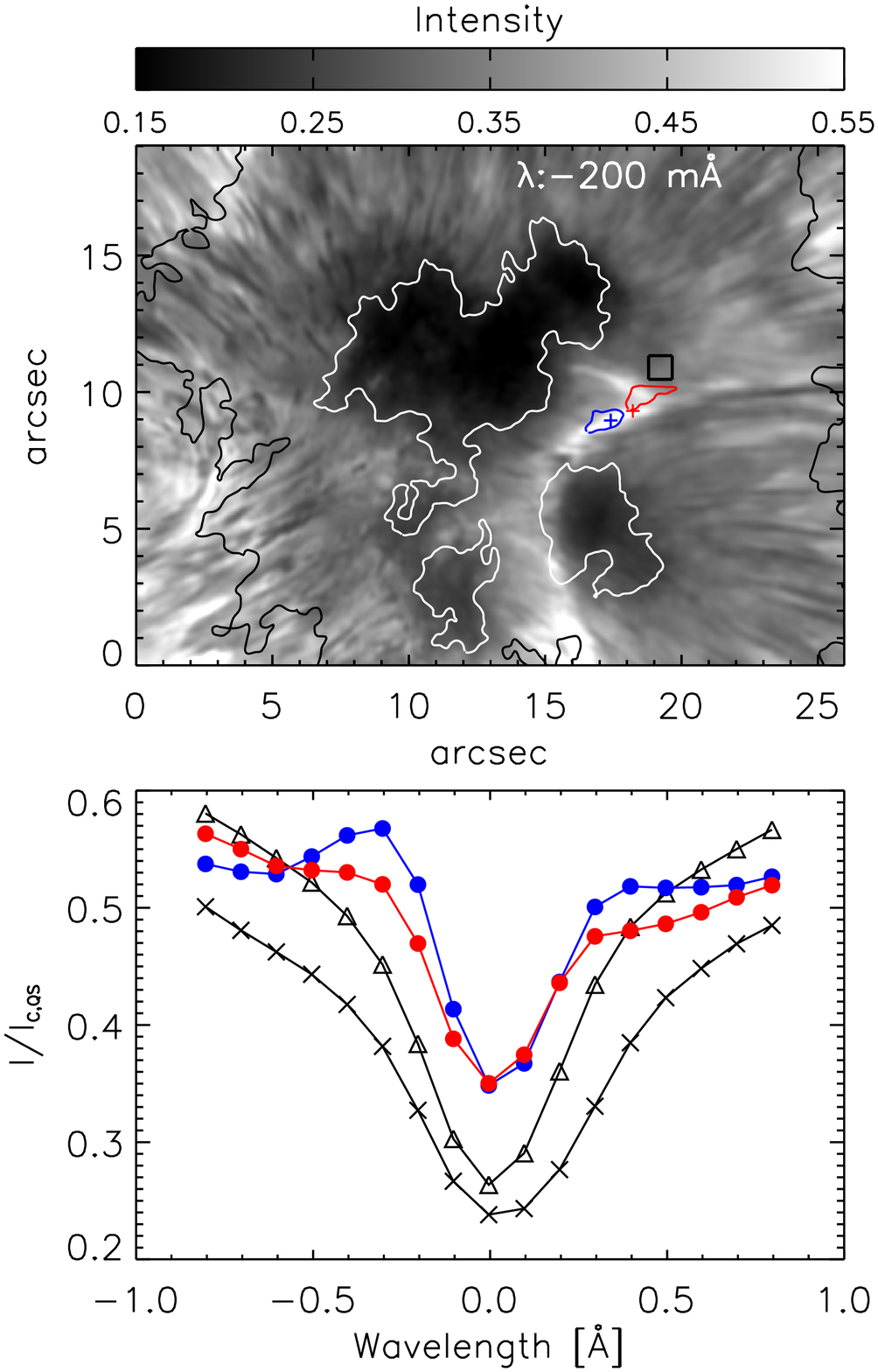} 
\vspace*{-10pt}
\caption{Intensity profile at the location of the brightening. Top
  panel: Ca filtergram at $-200$ m\AA\/ from line centre at minute~99.
  The blue and red contours indicate the photospheric
  blueshifts and redshifts. The blue and red plus symbols
  refer to the pixels whose intensity profiles are shown in
  Fig.~\ref{figure06}.  Bottom panel: mean Ca intensity profile in the
  quiet Sun (crosses), blue patch (blue
  circles), red patch (red circles) and thick
  black square close to the brightening (triangles). 
  The average profiles correspond to the scan at minute~99.}
\label{figure05}
\end{figure}

The sequence of field strength maps shows a small but discernible
increase of field strength with the onset of the blueshifts and
redshifts in the LB. This can be seen between rows 2 to 4 of
Fig.~\ref{figure03}. Interestingly, the field inclination bears a
strong association with the photospheric blue- and redshifts.  Rows 1
and 2 indicate that the red patch is an isolated concentration of
magnetic fields that are more horizontal than the surroundings.  The
field inclination increases in the red patch until it changes polarity
at minute~86 (black pixels in panel 4). The field also becomes more 
inclined in the blue patch as it evolves, but it never changes polarity.
There are no significant variations in the field azimuth related to
the evolution of either of the two patches.

Columns 6--11 of Fig.~\ref{figure03} indicate the temporal evolution
of the Ca intensity at different wavelength positions sampling the
region between the mid-photosphere to the low chromosphere. The Ca
filtergrams at $\pm$500 m\AA\/ exhibit isolated bright patches in
different parts of the FOV, which do not appear to be related to the
development of the photospheric flows.  The chromospheric response to
the redshifts and blueshifts is clearly seen at minute~93 (row 6) in
the form of the \textsf{V}-shaped brightening described earlier.  The
brightening starts between minute~83 and 90 (rows 4 and 5). It is
observed at all wavelength positions, but is relatively more intense in
the blue wing. In the line core (column 9), this feature is
comparatively subdued, although discernible. The strong chromospheric
brightening persists until the end of the observations.

The left panels of Fig.~\ref{figure04} show the mean value of various
parameters averaged within the blue patch as a function of time. The
mean velocity reaches a peak of 1.6 km~s$^{-1}$ nearly three minutes
after the emergence of the blueshifts in the LB. At this time, the
maximum velocity in the patch is 2.6 km~s$^{-1}$. For comparison, the
maximum velocity in the red patch is about 4.5~km~s$^{-1}$ at
minute~94, while the mean velocity is 2~km~s$^{-1}$ (not shown).  The
increase in velocity in the blue patch is accompanied by enhancements
of the field strength and inclination of 400 G and
25$^\circ$, respectively.  The mean value of the magnetic field and
inclination in the patch is 1.3~kG and 70$^\circ$ at the end of the
sequence. The red patch is some 500~G stronger, but the field strength
variation is similar.

The right panels of Fig.~\ref{figure04} depict the mean Ca intensity
at different wavelength positions.  The mean Ca intensity peaks around
minute~96, nearly 17~minutes after the blue patch emerges in the
photosphere.  All the wavelength positions exhibit an intensity
increase, although it is stronger in the blue wing compared
to the red wing. While the increase is gradual for the
spectral range from $-300$ to $+300$~m\AA, further out in the line it
is steeper (bottom right panel of Fig.~\ref{figure04}). The intensity
increases sharply from minute~94 to 96, particularly at $-300$ and
$-200$~m\AA, by 8\% and 6\%, respectively. In general, the mean blue
wing intensity exceeds the red wing intensity by about 5--10\%,
depending on the wavelength position.  Around the line core, the
intensity increases by some 15\% (top right panel) while in the
200--300 m\AA\/ range the increase is 20\% (middle right panel).
Finally, towards the photosphere (bottom right panel) the increase is
relatively weaker (6--10\%).

\subsection{Spectral characteristics of Ca brightening}
\label{chromo}
In this section, we describe the spectral characteristics of the
chromospheric brightening at the location of the photospheric
blueshifts and redshifts.  The top panel of Fig.~\ref{figure05} shows
a Ca filtergram at $-200$~m\AA\/ from line centre at minute~99. The
white and black contours enclose the umbra and penumbra of the
sunspot. The blue and red contours show the blue and red patches,
respectively. The bottom panel shows the average Ca intensity profile
emerging from them, as well as from the QS, at
minute~99. For comparison, the average profile in a
small region next to the brightening is also shown.

The \ion{Ca}{ii} 854.2~nm profile at the position
of the blueshifts shows a strong emission feature in the blue wing of
the line, around $-400$ and $-300$~m\AA. A similar feature can be
observed in the red wing, although it is less prominent.  Only a 
few pixels exhibit comparable emission peaks in the red patch.  The
average profile shows hints of emission in the blue wing and higher
intensities than in the red wing, but no strong peaks are observed.
Another important fact is that the line core is not Doppler shifted in
either of the two patches.  These \ion{Ca}{ii} 854.2~nm profiles are
rather different from those recorded outside the intense brightening
(triangles), which resemble the QS profile except for an overall
offset.  Profiles with asymmetric emission in the line wing have been
seen in observations \citep[e.g.,][]{2014ApJ...781..126O} and in
numerical simulations \citep{2009ApJ...694L.128L}.

\begin{figure*}[t]
\centerline{
\includegraphics[angle=0,width = 0.5\textwidth]{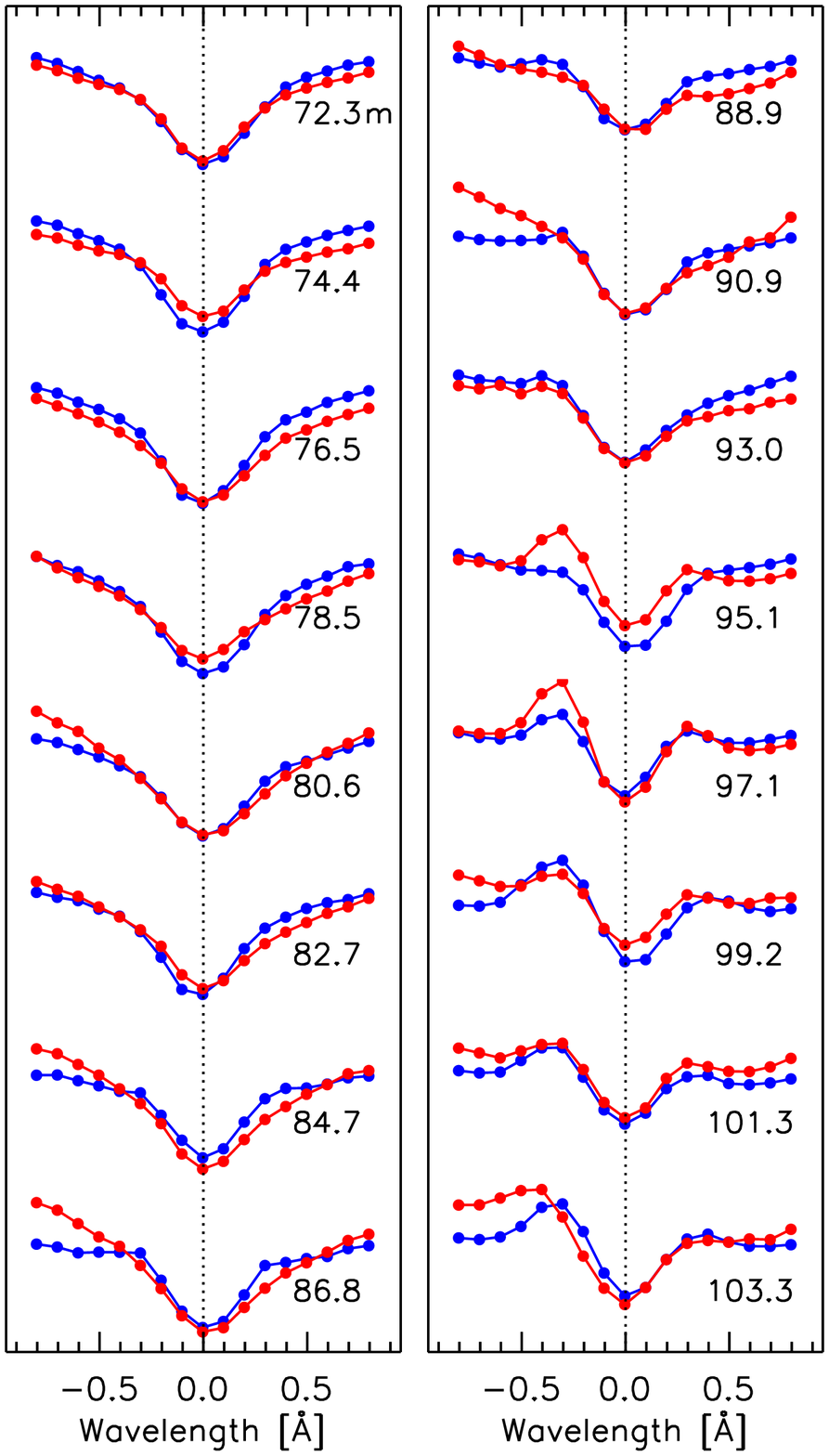}
\includegraphics[angle=0,width = 0.5\textwidth]{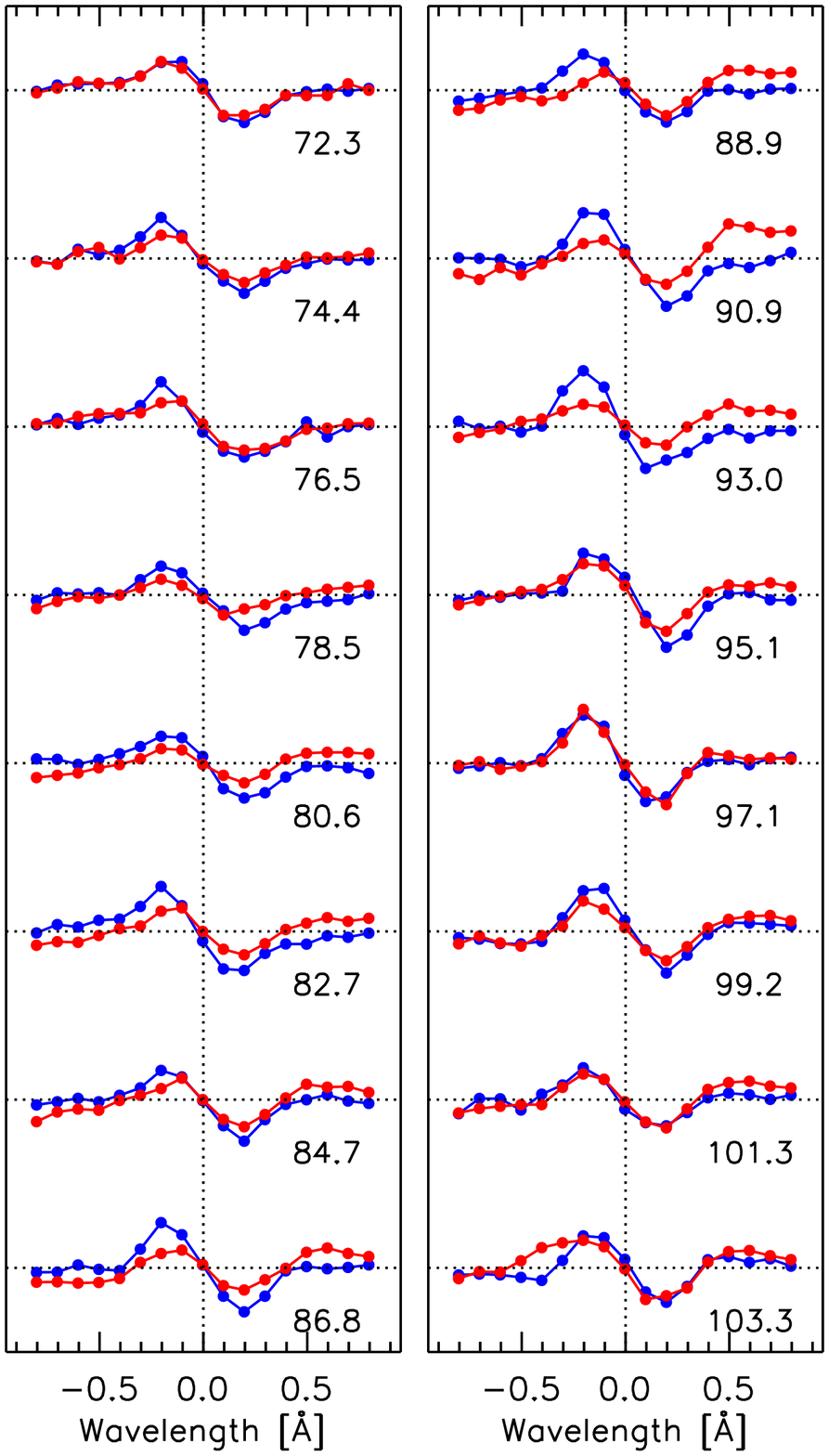}
}
\vspace{-55pt}
\caption{Temporal evolution of Ca~{\sc ii}~854.2 nm spectra in the
  blue and red patches (blue and red circles, respectively).
  Columns 1 and 2 correspond to Stokes $I$, while cols. 3 and 4 correspond to
  Stokes $V$.  Time increases from top to bottom and from left to
  right. The scaling of the $y$-axis in normalised intensity units is
  between $(0.2,0.7)$ and $(-0.05,0.05)$ for the two Stokes
  parameters. The profiles come from
  the pixel with the highest average intensity between $-300$ and
  $-400$~m\AA.}
\label{figure06}
\end{figure*}

Figure~\ref{figure06} shows the temporal evolution of the Stokes $I$
and $V$ profiles of the \ion{Ca}{ii} 854.2~nm line in the two patches.
We have selected the pixel with the highest average intensity between
$-300$ and $-400$~m\AA. Prior to the appearance of the blueshifts in
the photosphere, the Ca intensity profiles look normal. Around minute~84,
the profile from the blue patch starts to develop a knee in the blue
wing while the profile from the red patch remains unaffected. The
emission feature in the wings is seen around minute~89 and 95 in the two
patches, respectively. Both patches show asymmetric emission that
persists until the end of the sequence.
There are instances in which the emission in the red patch is stronger
than that in the blue patch (e.g. at minute~95 and 97).  As already
mentioned, the core of the line is not shifted significantly in any of
the patches.

The Stokes $V$ profiles show variations in asymmetry and amplitude
that seem to be unrelated to the development of flows in the
photosphere.  
In any case, the sign of the profiles remains unchanged
during the observations in both patches.

\subsection{Temperature and velocity in the chromosphere}
\label{cainv}
\subsubsection{Stokes inversions}
\label{inv}
We now examine the physical conditions in the chromosphere associated
with the brightenings that occur above the photospheric blueshifts and
redshifts. The conditions prior to the appearance of the brightenings
are also investigated to determine how the chromosphere is modified from
quiescence.

\begin{figure*}[t]
\centerline{
\includegraphics[angle=0,width = 0.43\textwidth]{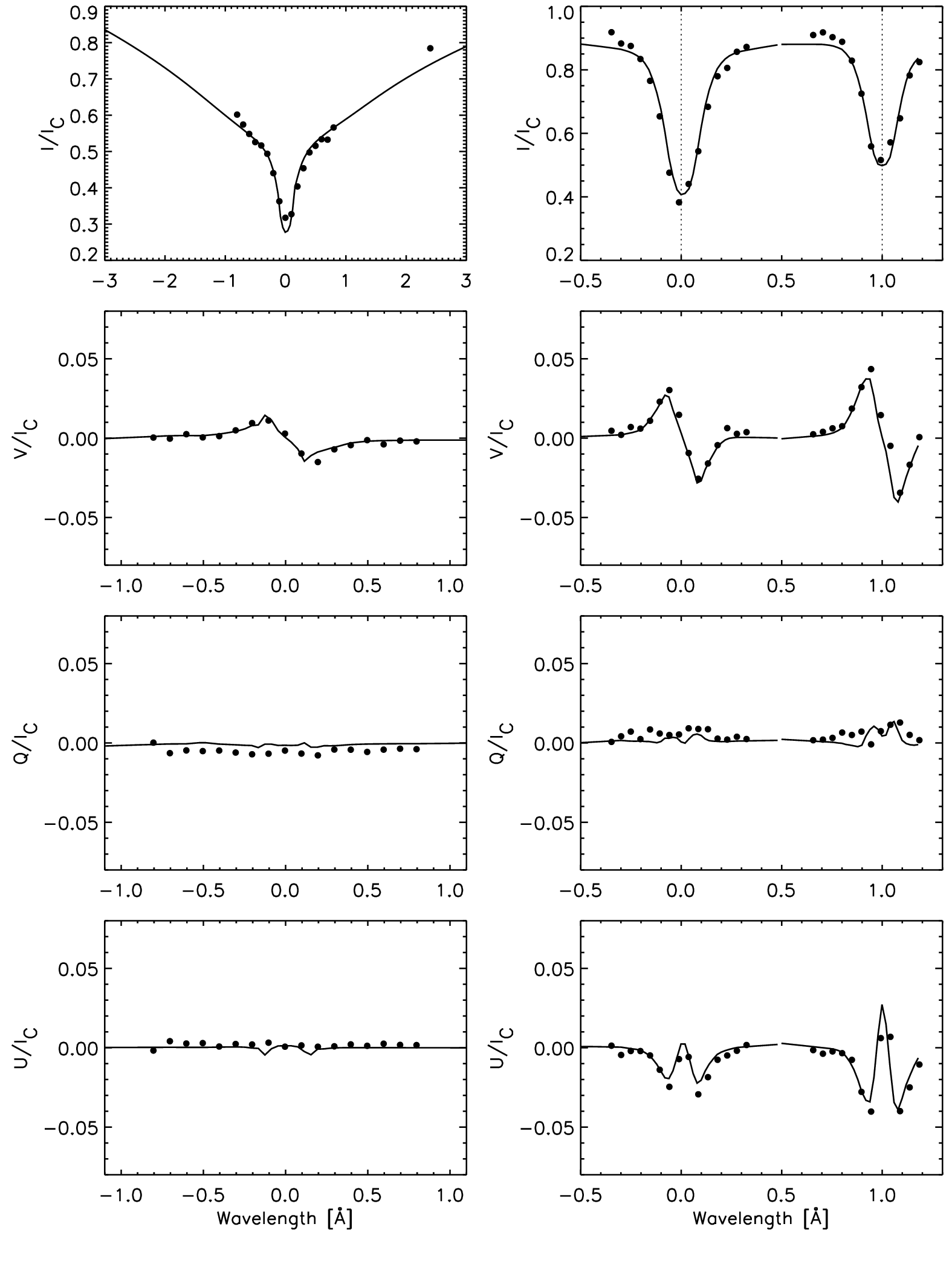}
\includegraphics[angle=0,width = 0.43\textwidth]{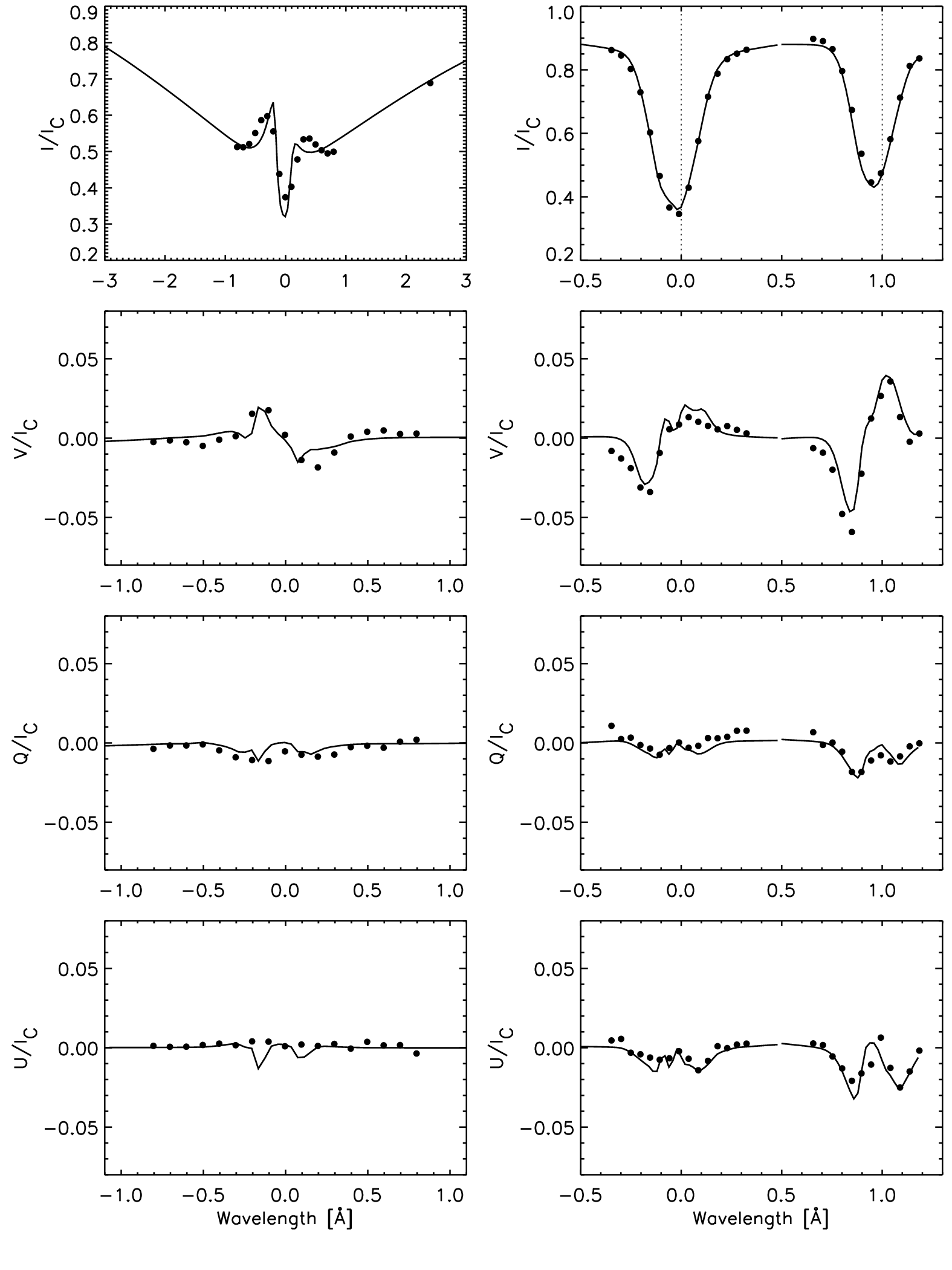}
}
\vspace{-10pt}
\caption{Results from NICOLE inversions. Columns 1 and 2: Stokes profiles
  of the Ca and Fe lines, respectively, for the case when there was no
  brightening in the LB. Columns 3 and 4: same as columns 1 and 2,
  except for the case when the brightening was observed. Filled circles
  and solid lines represent observed and synthetic spectra,
  respectively.}
\label{figure07}
\end{figure*}

In Fig.~\ref{figure07} we show the Stokes profiles observed in the
photosphere and the chromosphere before (columns 1 and 2) and during
the brightening (columns 3 and 4). The profiles come from the pixel in
the blue patch where the intensity between $-300$ and $-400$~m\AA~is 
highest (indicated with a cross in Fig.\ref{figure05}) at 16:19~UT
(minute~74) and 16:44~UT
(minute~99). They are representative of the typical
profiles encountered in this patch. As mentioned before, the Ca
intensity profile is normal during the quiescent phase and
subsequently exhibits an asymmetric emission in the line wing which is
the source of the observed brightenings.  The \ion{Ca}{ii} 854.2~nm
Stokes $V$ profiles do not change sign between the two instances of
time. The chromospheric $Q$ and $U$ signals are quite
noisy, irrespective of the presence or absence of the brightenings.
The \ion{Fe}{i} lines show significant changes,
particularly in Stokes $I$ and $V$. The $I$ profiles are at rest
during the quiescent phase and clearly blueshifted 25~minutes later.
Furthermore, the Stokes $V$ profiles, which initially have normal,
anti-symmetric lobes, change to highly asymmetric lobes with reversed
signs. The sign reversal is a projection effect due to the appearance
of highly inclined fields (see below).

To infer the temperature and velocity stratification in the
chromosphere, we employed the Non-LTE Inversion COde based on the
Lorien Engine \citep[NICOLE;][]{2015A&A...577A...7S}. The four Stokes
profiles of the \ion{Ca}{ii} 854.2~nm line were inverted along with
the two Fe lines (so as to cover a large height range and to provide
suitable boundary conditions for the NLTE calculations of
\ion{Ca}{ii} 854.2~nm). We ran NICOLE with two cycles,
allowing a maximum of 25 iterations per cycle.  The FALC model
\citep{1993ApJ...406..319F} served as an initial guess atmosphere, and
the parameters resulting from the first cycle were used to start the
second cycle. In the first cycle, temperature ($T$), LOS velocity
($V_{\rm LOS}$), LOS magnetic field ($B_{\rm LOS}$), and the two
transverse components of the magnetic field ($B_{\rm x}$, $B_{\rm y}$)
were perturbed with 2, 1, 2, and 1 nodes, respectively.  In the second
cycle, the number of nodes was changed to 6, 3, 3, and 2, plus 3 nodes
for microturbulence. To invert the profiles that show the Ca
brightening, the number of nodes in $T$ and $V_{\rm LOS}$ was
increased to 8 and 4, respectively. Stray light was included in the
inversion, assumed to be wavelength-independent.

The solid lines in Figure~\ref{figure07} show the
best-fit profiles resulting from the inversions. They match the
observed profiles quite well. With the exception of the continuum point,
the fit to the \ion{Ca}{ii} 854.2~nm intensity profile seen during the
quiescent phase is good.  Also the asymmetric emission in the line
wing is well reproduced, as is the nearly stationary line core during
the brightening (top panel in column 3). The observed and best-fit
Stokes $V$ profiles of the Ca line are in good
agreement. In the photosphere, the inversions reproduce the sign and,
to a great extent, the asymmetry of the Stokes $V$ signals associated
with the blueshifts.

\begin{figure}[!b]
\centering
\includegraphics[width = \columnwidth]{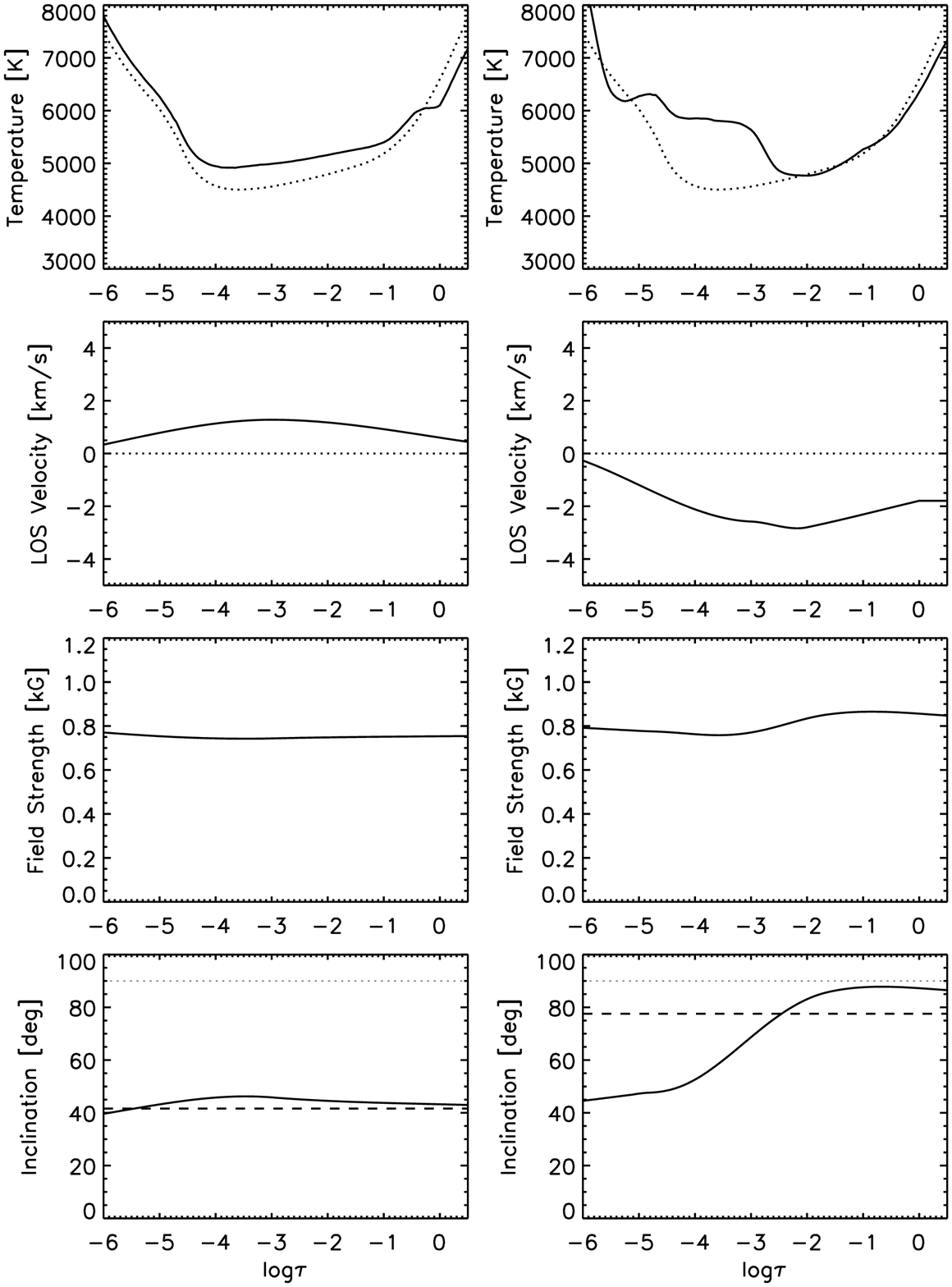}
\vspace{-25pt}
\caption{
Atmospheric parameters inferred from the NICOLE inversions
(solid lines). Left: quiescent phase. Right: brightening phase. 
The dashed lines show the values obtained from the SIR inversions. 
Dotted lines represent the FALC temperatures, zero velocities,
and horizontal fields in the respective panels.}
\label{figure07a}
\end{figure}

The model atmospheres returned by NICOLE are shown in
Fig.~\ref{figure07a}.
During the quiescent phase, the temperatures are slightly hotter than
those in the FALC model, except near the continuum forming layer.
There are redshifts of about 1.5~km~s$^{-1}$ at $\log\tau = -3.0$ and
weaker motions in higher and lower layers. The magnetic field strength
and inclination are nearly constant with height, reaching values of
about 750~G and 40$^\circ$, respectively.  At the time of the
brightenings, the temperature stratification exhibits a clear
enhancement between $\log\tau=-5.0$ and $-3.0$, with an excess of some
720~K relative to the quiescent phase.  The LOS velocity changes sign
and shows maximum blueshifts of 3~km~s$^{-1}$ at $\log\tau = -2.3$
that decrease to zero towards $\log\tau=-6.0$.  The
chromospheric field does not undergo significant changes from the
quiescent phase to the brightenings, contrary to the photospheric
field: in the photosphere, the field strength increases by about 100~G
and becomes nearly horizontal (with inclinations of approximately
88$^\circ$ to the local vertical). Above $\log\tau=-4.0$, i.e. beyond
the temperature minimum region, the field strength and inclination
continue to be 750~G and $40^{\circ}$ as in the quiescent phase.

To summarise, the blue patch in the LB is associated with rather
vertical fields at the beginning. Later the fields become highly
inclined in the photosphere, but do not change in the chromosphere.
The brightening over the patch is due to asymmetric emission in the
\ion{Ca}{ii} 854.2~nm line, caused by strong temperature enhancements
near the upper photospheric/lower chromospheric layers, together with
flows towards the observer of about 3~km~s$^{-1}$ in the mid-photosphere.

\subsubsection{Origin of the asymmetric Ca line-wing emission}
\label{syn}

The model atmosphere derived from the inversion of the observed
profiles reproduces the asymmetric emission that
causes the brightenings above the photospheric blueshifts. In this
section we further investigate the origin of such an asymmetry.

\begin{figure}[!t]
\centerline{
\includegraphics[angle=90,width = \columnwidth]{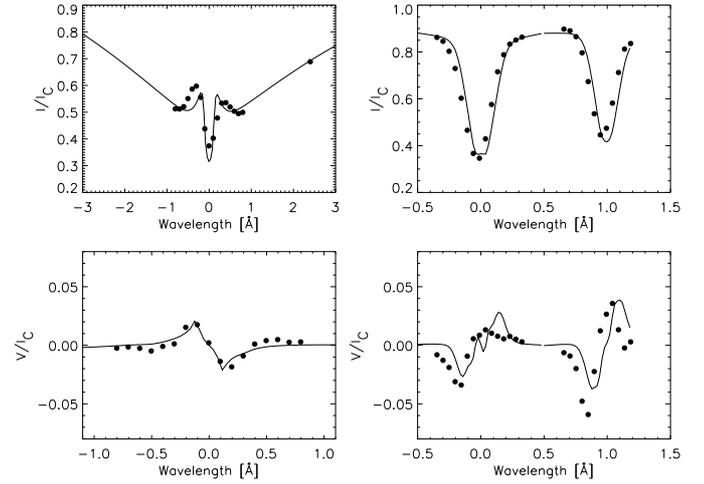}
}
\caption{Synthetic profiles (solid) emerging from the model atmosphere
  returned by NICOLE with zero velocities.  The other parameters
  remain unchanged. The observed profiles are shown with filled circles.}
\label{figure08}
\end{figure}

Temperatures and LOS velocities in high layers must play a more
important role in producing the asymmetric emission than the magnetic 
field since the latter does not change dramatically with respect
to the quiescent phase where the emission is absent. To assess the
influence of mass motions, we performed a spectral synthesis of the Ca
and Fe lines setting to zero all the velocities in the model
atmosphere returned by NICOLE. The result is shown in
Fig.~\ref{figure08}.  Without velocities, the asymmetric emission in
the Ca line disappears and we see symmetric emission peaks in both
line wings. The Stokes $V$ profiles remain more or less
unaffected, however, the lack of plasma motions shifts the Fe lines
away from their position. Moreover, the red and blue Stokes $V$
lobes become nearly antisymmetric, strongly deviating from the
observed profiles. Therefore, a gradient in the LOS velocity is needed
to account for the shape of the circular polarization profiles
recorded in the photosphere \citep{1975A&A....41..183I}.

This simple exercise demonstrates that (1) an enhancement in
temperature alone produces emission that is symmetric in both wings of
the Ca line, and (ii) the LOS velocity serves to enhance the emission
in one wing of the line, attenuating it in the other.  Thus, 
to generate the spectral signatures observed at the location of the
chromospheric brightenings, a combination of temperature and LOS
velocity perturbations is necessary.

\begin{figure}[t]
\centerline{
\includegraphics[angle=0,width = \columnwidth]{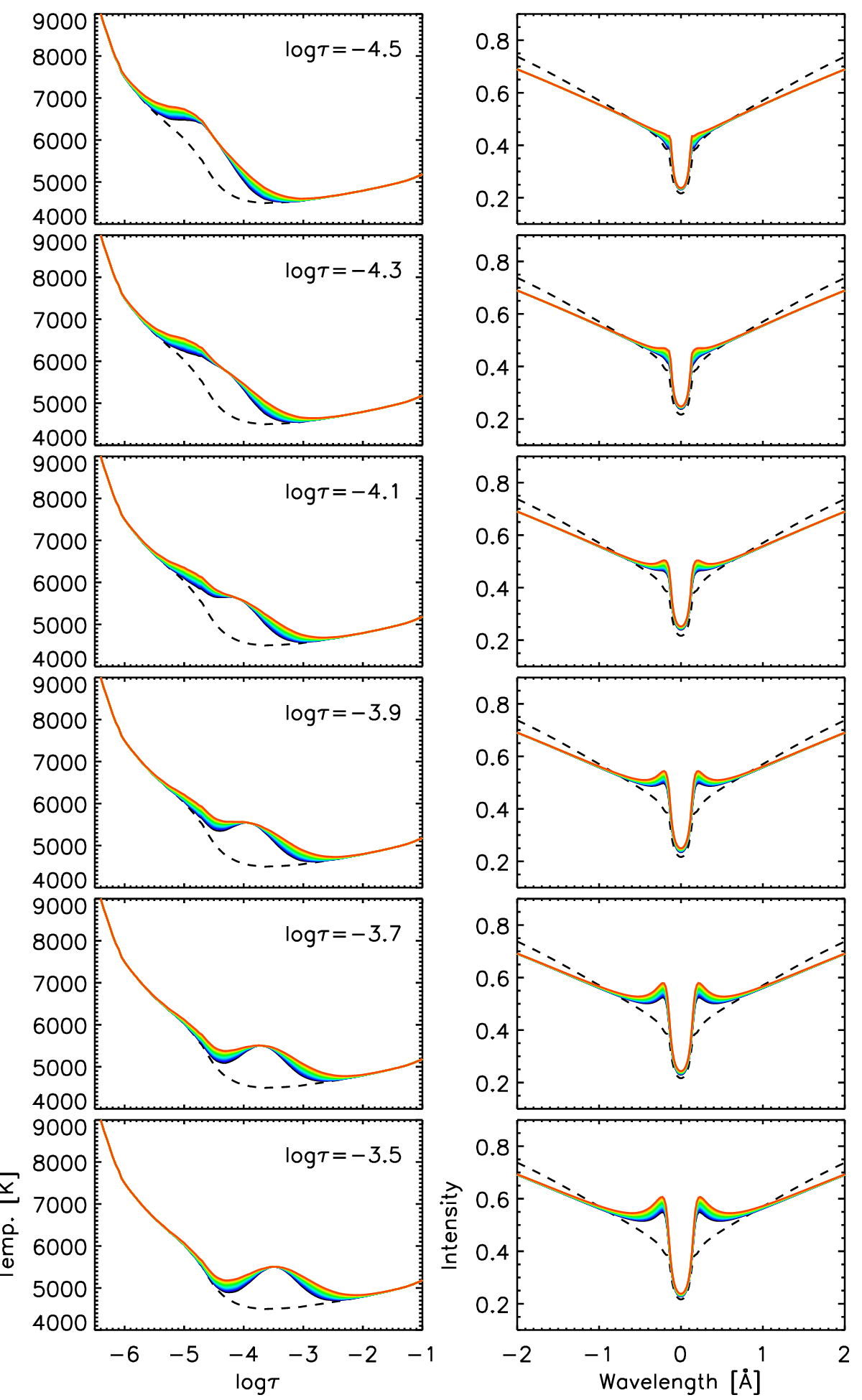}  
}
\caption{\ion{Ca}{ii} 854.2~nm profiles emerging from the FALC model
  with perturbed temperatures. The left panels show a
  Gaussian temperature enhancement of 1000~K at various optical
  depths. In each panel, the colours represent different widths of the
  perturbation with blue and orange corresponding to
  $\Delta \log \tau = 0.4$ and 0.6, respectively. The dashed
  line represents the unperturbed FALC atmosphere. The panels on the
  right are the corresponding intensity profiles calculated with
  NICOLE. The black dashed line shows the unperturbed profiles.}
\label{figure09}
\end{figure}

We now investigate the type of temperature enhancements 
required to produce symmetric line wing emission in the first
place. To this end, we synthesised the Ca line, modelling the
temperature enhancement as a Gaussian perturbation and changing its
width, position, and amplitude. We used the FALC model as the
reference atmosphere in these tests. The width of the temperature
perturbation, $\sigma$, was made to vary from 0.4--0.6 in units of
$\log\tau$, the position was set between -5.0$\le\log\tau\le$-3.0
and its amplitude was varied in the range 0--2000~K. The
temperature minimum is around $\log\tau=-3.8$ in the FALC model
\citep{2008A&A...480..515C}. For simplicity, all the velocities were
set to zero.

Figure~\ref{figure09} shows synthetic \ion{Ca}{ii} 854.2~nm line
profiles for temperature perturbations of $+$1000~K and
various widths located at different optical depths. From these
calculations we find that the strength of the emission depends on the
position of the perturbation: the deeper the perturbation below the
temperature minimum, the stronger the emission. This result is
somewhat expected because the coupling of the line source function to
the local temperature is usually weak (or non-existent) in the
chromosphere and strong in the photosphere
\citep[see][]{2013ApJ...764L..11D}.  
Another obvious effect of perturbations in deeper layers is that the
intensity ratio between the core and the wings becomes more
pronounced, because the intensity of the wings is greatly enhanced.

The width of the perturbation seems to affect the wing intensity close
to the knee-like feature of the line. Wider perturbations
increase the overall wing intensity, but the effect is significantly
smaller than that induced by the height of the perturbation.

As a final test, we perturbed the reference model using a linear
temperature increase with height. The results show that there is no
symmetric emissions in the line wing but the line exhibits a knee.
Moreover, such a perturbation enhances the overall spectral line
intensity. Increasing the temperature gradient also increases the
wing to core intensity ratio.

From these tests we conclude that the temperature perturbation needed
to produce emission in the wings of the \ion{Ca}{ii} line at the level
shown in Figure 9 (top left panel, solid line) cannot be located
much higher than $\log \tau \sim -4$ or much deeper than $\log \tau
\sim -2$.  That is, the brightening appears to be an upper
photospheric rather than a chromospheric phenomenon.  Additionally,
the temperature enhancement must extend over a minimum range of
optical depths to lead to measurable effects on the emergent profiles.
For a Gaussian perturbation of 1000~K located at $\log \tau = -3.5$,
the minimum width is $\Delta \log \tau = 0.6$.

\section{Discussion}
\label{discuss}

The photospheric LOS velocity maps obtained from the inversion of the
observed Stokes profiles reveal the formation of nearby patches of
upflowing and downflowing gas with opposite magnetic polarities in a
sunspot LB. Shortly after these patches are formed, we observe
chromospheric emission in the \ion{Ca}{ii}~854.2~nm line, which
suggests that these events are physically connected.  Here we try to
identify the photospheric processes responsible for the flows and then
interpret their subsequent coupling to the overlying chromosphere.

The first possibility that comes to mind is that the blueshifts and
redshifts represent the source and termination points of an Evershed
flow channel. The Evershed flow starts in bright penumbral grains,
which are blueshifted in the limb-side penumbra and have speeds of
about 0.5~km~s$^{-1}$ \citep{2006ApJ...646..593R}.  The blueshifts we
have observed are clearly stronger and, in addition, they coincide
with granular and not filamentary structures. Also, we do not see
clear proper motions of any of the patches, whereas it is known that
the termination points of the Evershed flow move radially outwards.

A scenario compatible with the observations is that of a small-scale
$\Omega$-loop that emerges in the LB and develops a siphon flow while
rising through the atmosphere. The blue- and redshifted patches
conform to a simple bipolar structure, which is magnetically connected
by a vector field pointing from the blueshifts to the redshifts. The
polarity of the magnetic field is positive (upwards) in the blue patch
and negative (downwards) in the red patch. A siphon flow should develop
along the loop because the red patch shows stronger fields
($\sim$400~G) and therefore the gas pressure is significantly
decreased there. This pressure imbalance between the footpoints causes
the gas to flow from the blue to the red patch.  The
loop footpoints do not separate during the sequence,
probably because the highly magnetised environment of the LB impedes
the horizontal expansion of the loop. The photospheric blueshifts and
redshifts are accompanied by a noticeable increase in field strength.
Moreover, the blue patch coincides with an emerging, elongated
granule in the LB. Taken together, this suggests that we
have witnessed a small-scale flux emergence event in the LB.

\begin{figure}[t]
\centerline{
\includegraphics[angle=90,width = \columnwidth]{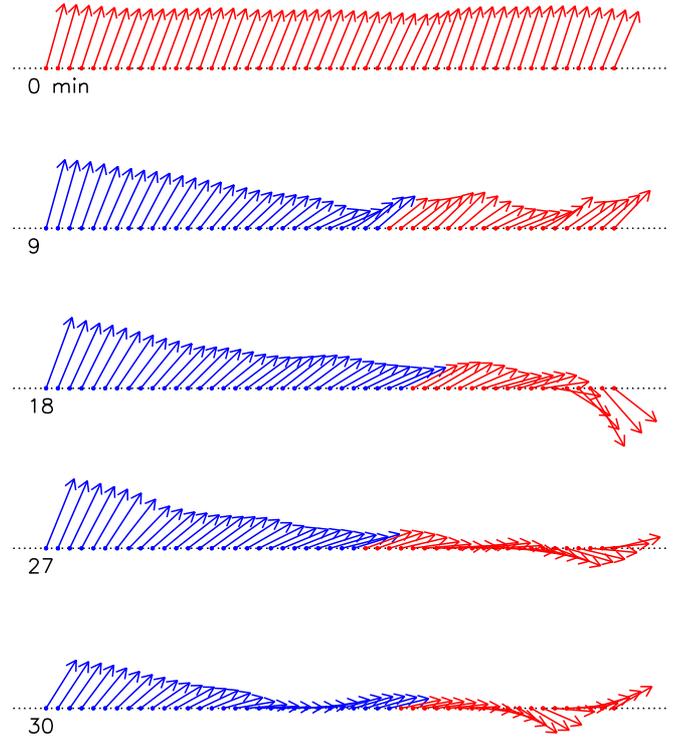}
}
\caption{Sketch of the flat $\Omega$-loop emerging in the light
  bridge.  The arrows have been drawn for a cut passing along the blue
  and redshifted patches. The length and orientation of the arrows are
  proportional to the field strength and inclination inferred from the
  inversion. The start time corresponds to minute~72 of the time
  sequence.}
\label{figure10}
\end{figure}

According to our results, the magnetic field in the footpoints is
rather inclined and therefore the $\Omega$-loop must have a
flat magnetic field topology. Figure~\ref{figure10} shows a
sketch of the magnetic field orientation along the loop for some of
the time steps. As the loop evolves, the field remains dominantly
horizontal, only changing polarity in the redshifted footpoint.  This
is significantly different from quiet-Sun flux emergence, where the
field becomes more vertical as the loop rises in the atmosphere
\citep{2009ApJ...700.1391M}.  Light bridges are known to be structures 
where the umbra is subject to convective instabilities
\citep{2006A&A...453.1079J, 2012ApJ...755...16L}. The granular
morphology of the LB thus offers a natural location for the emergence
of small-scale magnetic fields, aided by the upflows of convective
cells. The emerging fields are highly inclined or even horizontal,
suggesting that the strong sunspot fields around and
above the loop influence the way in which they emerge into the LB.
This could explain the asynchronous appearance of the footpoints of
the loop.

The magnetic fields associated with the loop continue their ascent
into the chromosphere where strong brightenings are observed 17~minutes
after the development of the blueshifts. Compared with earlier LB
studies based on \ion{Ca}{ii}~H broadband imaging, we have the entire
\ion{Ca}{ii}~$8542$ profile. The emission is located in the blue wing
of the line, without an obvious Doppler shift of the line core. Light bridges
are known to exhibit dynamic chromospheric phenomena as described in
Sect.~\ref{intro}, with magnetic reconnection often cited as
their driving mechanism \citep{2003ApJ...589L.117B,
2006A&A...453.1079J, 2009ApJ...704L..29L,2009ApJ...696L..66S}. In
this case, magnetic reconnection is a plausible explanation for the
observed brightening because the sunspot magnetic field over the LB
runs across the redshifted patch, which should be favourable for
component reconnection between the emerging field and the sunspot
field.

We find that large enhancements in temperature and a moderate LOS
velocity towards the blue, particularly in the region between
the upper photosphere and the lower chromosphere (i.e., the minimum
temperature region), are sufficient to produce the spectral features
seen in \ion{Ca}{ii} 854.2~nm above the photospheric blueshifts. This
layer of the solar atmosphere is extremely sensitive to fluctuations
in the thermodynamic and kinematic parameters, all of which are
directly or indirectly influenced by the magnetic field. The magnetic
loop first emerges in the weak-field environment of the LB, but
then it encounters the strong magnetic fields of the
sunspot during its ascent, oriented at an angle with respect to the
loop.  These gradients in the magnetic field could lead to
temperature enhancements through ohmic dissipation of
electric currents \citep{2003Natur.425..692S,2013A&A...555A.123B}.
We observe a gradual increase in the blue wing intensity close to the
knee of the spectral line, followed by a 
steep rise. The steep rise would be compatible with the
idea of magnetic reconnection where the free energy is impulsively
released.

The \ion{Ca}{ii} line core is unshifted, suggesting
that the rising loop does not reach the chromosphere. In fact, the
chromospheric magnetic field remains largely unaffected, unlike the
photospheric field which becomes nearly horizontal during the
emergence of the loop. Therefore, the emerging magnetic flux is
confined to or trapped in the upper photospheric layers, possibly
stopped by the overlying magnetic field of the sunspot. However, we do
not rule out the possibility that the emerging magnetic field
eventually reaches chromospheric heights, since the observing sequence
ended before the flow structures disappeared.

The profiles analysed in the present study are not very different from
those observed in flux emergence regions outside of sunspots
\citep{2014ApJ...781..126O}.  \citet{2015arXiv150303846D} explained a
similar (asymmetric) emission but in the red wing of the 
\ion{Ca}{ii} 854.2~nm line, within a flux-emerging region. In their
observations and MHD simulations, the presence of a velocity field
gradient shifts the absorption profile in the upper photosphere to
shorter wavelengths. This allows photons to escape more
easily in the red line wing, thus producing enhanced emission in
that wing and the opposite effect in the other. The main difference
here is that, although we also have upflowing material, the emission
occurs in the blue line wing. Therefore, in our data the emission is
obviously not caused by the same opacity effect
reported by \citet{2015arXiv150303846D}, although the velocity field
seems to be responsible for the observed asymmetry.  There are also
similarities with the Class-1 chromospheric transient events reported
by \citet{2013ApJ...779..143R} and \citet{2014drews}.  These events
also exhibit a stationary line core and impulsive intensity increases
that occur within a span of 3-4 minutes and are compatible with the
sharp rise detected around minute~94.

In summary, we have interpreted the development of photospheric blue-
and redshifts in a sunspot LB as the emergence of a small-scale, flat
$\Omega$-loop, whose topology and evolution is influenced by the
strongly magnetised environment of the sunspot.  The Doppler shifts
occurring in the two footpoints can be explained naturally in terms of
a siphon flow, which develops as a consequence of a significant field
strength imbalance.  The ascending loop is probably trapped by the
overlying sunspot field and the subsequent interaction is responsible
for boosting the temperature near the upper photospheric/lower
chromospheric layers, producing peculiar asymmetric emission in the
blue wing of the \ion{Ca}{ii} 854.2~nm line.  Since we do not know 
the ultimate fate of this emission, the physical mechanism behind it
remains an open question. By combining multiple spectral lines, which
cover a large height range from the photosphere to the chromosphere,
it is possible to determine the coupling between the different layers
of the atmosphere. This will be the topic of further investigation.

\begin{acknowledgements}
  Part of this work was done while R. E. L was a Visiting Scientist at
  the Instituto de Astrof\'isica de Andaluc\'ia (CSIC), where he
  continued the analysis initiated at the Udaipur Solar Observatory as
  a Post Doctoral Fellow.  Financial support by the German Science
  Foundation under grant DE 787/3-1, the European Commission's
  FP7 Capacities Programme under Grant Agreement number 312495
  (SOLARNET project), and by the Spanish Ministerio de Econom\'{\i}a y
  Competitividad through projects AYA2012-39636-C06-05 and
  ESP2013-47349-C6-1-R (including a percentage from European FEDER
  funds) is gratefully acknowledged.  R. E. L is thankful to SOLARNET 
  for funding the visit to the Instituto de Astrof\'isica de Andaluc\'ia 
  (CSIC) under the Mobility Programme for Young Researchers. The Swedish 1-m Solar Telescope
  is operated by the Institute for Solar Physics of Stockholm
  University in the Spanish Observatorio del Roque de los Muchachos of
  the Instituto de Astrof{\'\i}sica de Canarias. We thank the referee for the
  useful suggestions and comments. This research has
  made use of NASA's Astrophysical Data System.
\end{acknowledgements}

\bibliographystyle{aa} 
\bibliography{louis_reference_v10}

\end{document}